\documentstyle[preprint,tighten,eqsecnum,aps]{revtex} 
\begin{document}
\draft
\preprint{UTS-DFT-94-08}
\title{Conformal p-branes as a Source of Structure in Spacetime}
\author{A.Aurilia}
\address{
Department of Physics, California State Polytechnic University,\\
Pomona,
CA 91768}

\author{A.Smailagic}
\address{
International Center for Theoretical Physics,\\ Strada Costiera 11,
34014 Trieste, Italy}

\author{E.Spallucci}
\address{
Dipartimento di Fisica Teorica Universit\`a di Trieste,\\ INFN
Sezione
di Trieste,\\  Strada Costiera 11, 34014 Trieste, Italy}
\maketitle
\begin{abstract}
We discuss a model of a conformal p-brane interacting
with the world volume metric and connection. The purpose of the
model is to suggest a mechanism by
which gravity coupled to p-branes leads to the formation of
{\it structure} rather than homogeneity in spacetime.
Furthermore, we show that the formation of structure is accompanied
by the appearance of a {\it multivalued} cosmological constant,
i.e., one which may take on different values in different domains,
or {\it cells}, of spacetime. The above results apply to a broad
class of non linear gravitational lagrangians as long as metric and
connection on the p-brane manifold are treated as independent
variables.
\end{abstract}
\bigskip

\section{Introduction}

This paper has two main objectives. The first is to discuss a new
effect that the dynamics of extended objects (p--branes) may have on
the geometry of spacetime. The second, allied, objective is to
introduce a class of gravity theories in (p+1)--dimensions
characterized by the formation of structure on the p--brane
manifold.

To elaborate
further on these points, and partly to motivate our work, we recall
one of the basic tenets of General Relativity, namely, that the
matter content of the universe shapes the spacetime geometry and,
conversely, that the geometry ``~guides~'' the motion of material
{\it particles} along geodesic lines. The notion of ``~geodesic
motion~'' can be generalized to incorporate the world--track swept by {\it
extended objects}, say strings and membranes, in curved spacetime
\cite{noi}.
However, once this is done, one finds that a new possibility arises
from the interplay between geometry and dynamics, namely, the {\it
formation of structure in spacetime.} This structure consists of
separate vacuum domains, or {\it cells} of spacetime, each
characterized by
a distinct geometric phase; that is, the background geometry could
be Riemannian or Minkowskian in one domain, of Weyl type in another
or Riemann--Cartan in yet another cell, etc., with the highest
concentration of matter to be found on the domain walls separating
each cell. The interior of each cell constitutes a ``~false
vacuum~'' to the extent that it is characterized by a distinct value
of the cosmological constant or vacuum energy density. Since each
cell has a dynamics of its own, this overall cellular structure is
an ever changing one, and the qualitative picture that comes to our
mind is that of a ``~frothiness~'' in the very fabric of spacetime.

In different guises and with different objectives in mind, a
cellular structure in spacetime has been invoked before \cite{cole},
and is implicitly assumed,
for instance, in connection with the idea of chaotic inflation
\cite{linde}, or in connection with the geometrodynamic idea of {\it
spacetime foam} as the inevitable consequence of quantum
fluctuations in the gravitational field \cite{wheeler}. However,
since very little is known about quantum gravity, a mathematical
implementation of these ideas has always been exceedingly difficult,
or vague. {\it In contrast, the aim of this paper is to show that
one can ``~tunnel through the barrier of ignorance~'' about quantum
gravitational effects and discuss the formation of
a multiphase cellular structure in spacetime as a consequence of the
classical dynamics of p--branes coupled to gravity.} This brings us
to our second and more detailed objective, i.e., the introduction of
a new class of gravity theories in (p+1)--dimensions. To trace the
genesis of this approach we recall that the notion
of a spacetime with many geometric phases
originated in an earlier attempt to deal with the phenomena of
vacuum decay and inflation \cite{lego}, and, more recently, from a
stochastic approach to the
dynamics of a string network in which we have shown that domains of
spacetime (voids) characterized by a Riemannian geometry and a
nearly uniform string distribution, appear to be separated by domain
walls characterized by a Weyl type geometry and by a discontinuity
in the string distribution \cite{vanz}. In this approach, based on a
stochastic interpretation of the Nambu--Goto action,
the geometry of spacetime
is not preassigned, but it is required to be compatible with the
matter--string distribution in the universe. Thus, the string
degrees of freedom are coupled to both metric $\gamma_{mn}$ and
connection $\Gamma^\l{}_{mn}$ of the {\it ambient spacetime} through
the curvature scalar $R=\gamma^{mn}R_{mn}(\Gamma)$. The general
philosophy of this paper is the same, i.e., geometry and matter
distribution must be self consistent and not preordained. In this
paper, however, by ``~geometry of spacetime~'' we mean the {\it intrinsic}
geometry of the Lorentzian p--brane manifold and not the geometry of
the target space in which the p--brane is imbedded and which we
assume, for simplicity, to be a $D$--dimensional Minkowski
spacetime. From this vantage point, the p--brane {\it classical}
action that we suggest below in Eqs. (2.1 and 2.5), can be
interpreted as the action for gravity in (p+1)--dimensions coupled
to some ``~scalar fields~'' represented in the action by the
imbedding functions with support on the p--brane manifold.
The payoff of this particular choice of action is the possibility,
not usually contemplated by
conventional General Relativity, of a multiphase intrinsic geometry
that may form on a p--brane manifold. In fact, the main result of
this paper is a mechanism, {\it coded in the condition} (2.14), by
which gravity coupled to {\it extended
objects} manages to produce structure rather than uniformity in
spacetime. The source of this mechanism can be traced back to two
key properties of our model: the first property is that the
gravitational term in the action is described by an {it analytic}
function of the scalar curvature on the p--brane manifold; the
second property is that the energy--momentum tensor of the p--brane
is {\it traceless}.These properties are coded in the two terms of
the action (2.5): the first property (analyticity of the
gravitational term) is simply {\it assumed}, with no other
justification tha to serve our purpose, which is to arrive at the
condition (2.14) bypassing quantum gravitational effects; the second
property (tracelessness of the energy--momentum tensor) is {\it
enforced} by restricting our consideration to {\it conformal p--
branes} defined by Eq.(2.1). The rationale for this choice of p--
brane action is that any other choice would result in the appearance
of the trace of the energy--momentum tensor on the right hand side
of Eq.(2.14), thereby invalidating our conclusions.

The main body of the paper, section II, is divided into three
subsections. In subsection A, we introduce the action
functional for the
conformal p--brane non--minimally coupled to the world volume
metric and connection, which we consider as independent variables.
In subsection B, we describe the solution of the classical field
equations corresponding to a Riemannian geometry over the p--brane
world volume.
In subsection C, we show how the same classical field equations
admit another type of solution. For a generic p--brane, with $p>1$, this
solution corresponds to a Riemann--Cartan geometry
characterized by a traceless torsion tensor. The string case is
exceptional in that the solution of the field equations corresponds
to a Weyl geometry.

\section{The Action}

\subsection{Classical p--brane dynamics}
	In the conventional approach originated by Dirac, Nambu and
Goto, p--branes are treated as $(p+1)$--dimensional manifolds {\it
imbedded} in a $D$--dimensional spacetime. Alternatively, one may elect to
focus on the {\it intrinsic geometry} of the p--brane manifold,
regardless of the imbedding in the ambient spacetime. Our action
integral reflects both points of view. A first step toward this
``~hybrid~'' model was suggested for the string
by Howe and Tucker \cite{ht}. On purely
dimensional grounds, the Howe--Tucker string action, which is
equivalent to that of Nambu and Goto, is invariant
under Weyl rescaling of the world metric $\gamma_{mn}$ and, as a
consequence, the string classical energy--momentum tensor has
vanishing trace. {\it This is the key property of strings which we
wish to extend to a generic p--brane}.As anticipated in the Introduction,
one way to achieve this is to give up the world--volume
interpretation of the action and to formulate p--brane dynamics in a
manifestly Weyl invariant form. The extension of the Howe--Tucker
action, though feasible, does {\it not} meet this requirement
\cite{suga}. Rather,
the Weyl invariant classical action for a p--brane is \cite{Duff}
\begin{equation}
S_{\rm C}=-\kappa\int_W d^{p+1}\xi\sqrt{-\gamma}\left[
{1\over(p+1)}\gamma^{mn}\partial_m X^\mu \partial_n X_\mu
\right]^{\left(p+1\right)/2}\ ,\label{uno}
\end{equation}
where $\kappa$ is the p--brane surface tension, $\xi^m$,
$m=0,\dots,p$ denote
the world volume coordinates with world volume metric $\gamma_{mn}$, and
$X^\mu$, $\mu=0,\dots, D-1$, denote spacetime coordinates with a
flat metric $\eta_{\mu\nu}$.
Since the combination
$\displaystyle{\sqrt{-\gamma}\left(\gamma^{mn}\right)^{(p+1)/2}}$ is
Weyl invariant for any $p$, the p--brane energy momentum tensor
\begin{eqnarray}
T_{mn}&\equiv& -{2\over\sqrt{-\gamma}}{\delta
S_C\over\delta\gamma^{mn}}
\nonumber\\
&=&\kappa \partial_m X^\rho \partial_n X_\rho\left[
{1\over(p+1)}\gamma^{pq}\partial_p X^\mu \partial_q X_\mu
\right]^{\left(p-1\right)/2}+\nonumber\\
&-&\kappa\gamma_{mn}\left[
{1\over(p+1)}\gamma^{pq}\partial_p X^\mu \partial_q X_\mu
\right]^{\left(p+1\right)/2} \label{timunu}
\end{eqnarray}
is {\it traceless}, i.e. $T^m{}_m=0$.

The next step in our approach is to add to the action an explicit
symmetry breaking
term which accounts for the ``~intrinsic~'' gravitational
interaction on the p--brane manifold. Note that a generic term of
this type is expected to arise in the effective action as a
consequence of quantum corrections \cite{smail}. However, for our
specific purposes, stated in the Introduction, we define on the p--
brane manifold a world hypersurface {\it
affine connection} $\Gamma^s{}_{mn}$,
$(s,m,n=0,\dots,p)$ through
an interaction term $L_{int.}(R)$ which is assumed to be an {\it
analytic}, but otherwise arbitrary
function of the world volume scalar curvature $R$.
We do not select the usual Christoffel connection because
this choice would impose a Riemannian geometry on the
world volume. Instead, as explained in the Introduction, we consider
the conformal p--brane geometry
as a dynamical quantity to be determined by the equations of motion.
In this general case, the strength of the connection is measured by
the curvature tensor

\begin{equation}
R^l{}_{mns}\equiv \partial_n\Gamma^l{}_{ms}-
\partial_s\Gamma^l{}_{mn}
+\Gamma^l{}_{an}\Gamma^a{}_{ms}-\Gamma^l{}_{as}\Gamma^a{}_{mn}\ ,
\label{due}
\end{equation}
and the corresponding contracted curvature tensor and curvature
scalar, are given by
\begin{equation}
R_{ms}(\Gamma)=R^l{}_{mls}\ ,\qquad
R(\gamma,\Gamma)=\gamma^{ms}R_{ms}\ .
\label{treuno}
\end{equation}
Note that $R_{ms}$ does {\it not} depend on the world volume metric,
but is a function of the connection alone. Futhermore, $\gamma^{mn}$
projects out the symmetric part of $R_{mn}$ in the definition of the scalar
curvature $R$. Against this background, the action describing our
model is

\begin{equation}
S(X,\gamma,\Gamma)=S_{\rm C}(X,\gamma)+
\int_W d^{p+1}\xi\sqrt{-\gamma}L_{int.}(R)
\label{tre}
\end{equation}
in which $L_{int.}(R)$ can be regarded either as an assigned
function of $R$, or
as a generic analytic function to be determined by the equations of
motion. Note that in the action (\ref{uno}), the p--brane is {\it
minimally} coupled to the world volume metric in a Weyl invariant
manner, whereas in the action
(\ref{tre}) we have introduced a non--minimal interaction term.
Except for a special form of $L_{int.}(R)$, to be discussed shortly,
it is to be expected that this term breaks the conformal symmetry of
the action (2.1) and our immediate objective is to discuss the main
dynamical consequence of this symmetry breaking term, namely, the
formation of structure accompanied by the appearance of a
multivalued cosmological constant on the p--brane manifold.

Varying eq.\ (\ref{tre}) with respect to the p--brane coordinates
$X^\mu$ , we find
\begin{equation}
\partial_m\left(\sqrt{-\gamma}\gamma^{mn}\partial_n X^\mu\right)=0\ .
\label{quattro}
\end{equation}
Equation\ (\ref{quattro}) is the ``~free~'' wave equation for the
p--brane field $X^\mu(\xi)$ and would represent the whole content
of our model in the absence of the intrinsic gravitational term. Eq (2.6) is
essentially a generally covariant Klein--Gordon equation
with respect to the world volume metric $\gamma_{mn}$, and does not
depend on the connection $\Gamma$. As a matter of fact, $X^\mu(\xi)$
behaves as a scalar multiplet under a general coordinate transformation
$\xi^m\rightarrow \xi^{'m}=
\xi^{'m}(\xi)$, and, therefore, general covariance only determines
the coupling to the metric.

Next, varying eq.\ (\ref{tre}) with respect to the world volume
metric, we find

\begin{eqnarray}
&&\gamma^{mn}\left[
{1\over(p+1)}\gamma^{pq}\partial_p X^\mu \partial_q X_\mu
\right]^{\left(p+1\right)/2}-\gamma^{mi}\gamma^{nj}\partial_i
X^\rho\partial_j X_\rho
\left[
{1\over(p+1)}\gamma^{pq}\partial_p X^\mu \partial_q X_\mu
\right]^{\left(p-1\right)/2}+\nonumber\\
&&-L_{int.}^\prime (R)R^{(mn)}(\Gamma) +{1\over
2}L_{int.}(R)\gamma^{mn}=0\ ,
\label{cinque}
\end{eqnarray}
where the prime denotes derivation with respect to $R$, $R_{(mn)}$
is the symmetric part of the contracted curvature tensor, and $\nabla_a$ is
the covariant derivative with respect to the $\Gamma$ connection.
In the absence of non--minimal interactions, eq.\ (\ref{cinque})
reduces to a relationship between the world volume metric
$\gamma_{mn}$ and the induced metric
$g_{mn}=\partial_m X^\mu \partial_n X_\mu$, modulo an arbitrary Weyl
rescaling.
This relationship is changed by $L_{int.}(R)$
and eq.(\ref{cinque}) encodes the coupling between the p--brane
field, metric and connection in the general case.

Finally, we have to vary  the action with respect to the connection.
In order to do this, it may be useful to recall the formula
\begin{equation}
\gamma^{ms}\delta_\Gamma R_{ms}(\Gamma)=\gamma^{ms}\left[\nabla_l
\delta\Gamma^l{}_{ms}-
\nabla_s \delta\Gamma^l{}_{ml}\right]\ .
\label{ocinque}
\end{equation}
Hence, the requirement
\begin{equation}
\delta_\Gamma S=\int_W d^{p+1}\xi\sqrt{-\gamma}L_{int.}^\prime (R)
\gamma^{ms}\delta_\Gamma R_{ms}(\Gamma)=0
\label{osei}
\end{equation}
gives, after an integration by parts:
\begin{equation}
\nabla_l\left[\sqrt{-\gamma}L_{int.}^\prime (R)
\gamma^{mn}\right]-\nabla_s\left[\sqrt{-\gamma}L_{int.}^\prime
(R)\gamma^{ms}
\right]\delta^n_l=0\ .
\label{osette}
\end{equation}
Taking the trace over the pair $(l,n)$, we find that
$\displaystyle{\nabla_n(\sqrt{-\gamma}L_{int.}^\prime
(R)\gamma^{mn})=0}$,
so that we can write eq.\ (\ref{osette}) in the form
\begin{equation}
\nabla_l\left[L_{int.}^\prime (R)\sqrt{-\gamma}\gamma^{mn}\right]=0\ .
\label{sei}
\end{equation}
Equation (\ref{sei})
relates $\Gamma^m{}_{nr}$ to $\gamma_{mn}$ and can be used to
determine the world volume geometry.
In order to see this, we note that the first two
terms in eq.(2.7) represent just the traceless p--brane energy--
momentum tensor (\ref{timunu}). Therefore, if we take the trace of eq.\
(\ref{cinque}), the dependence
on $X^\mu(\xi)$ disappears and we obtain the following relation
between the metric and the connection,
\begin{equation}
RL_{int.}^\prime (R)-{p+1\over 2}L_{int.}(R)=0 .\label{sette}
\end{equation}
Equation (\ref{sette}) was first derived in ref. \cite{to} as a
condition on a broad class of non--linear gravitational lagrangians
leading to the same Einstein equations obtained from the usual
Hilbert action. Volovich \cite{vol} has subsequently applied that
condition to the case of gravity on the world--sheet of a string and
our work was largely inspired by these papers. Regarding
equation (2.12), essentially one has two options: the first is
to interpret eq.(\ref{sette}) as a {\it differential} equation for
$L_{int.}$, in which case the solution is easily found to be
\begin{equation}
L_{int.}(R)={\rm const.}\times R^{(p+1)/2}\ . \label{otto}
\end{equation}
This function is analytic and invariant under Weyl rescaling. Thus,
for any extended object, there is a non--minimal gravitational
coupling which is singled out by the Weyl invariance of the action.
However, in general one starts from an {\it assigned}, non--invariant
interaction Lagrangian,
so that the form of eq. (\ref{sette}) is fixed {\it a priori}. This
is our second option. As an example, if we specialize the model to
the {\it bag} case, $p=3$, a suggestive form of $L_{int.}(R)$ is:
$\displaystyle{
L_{int.}(R)=\rho -\mu^2 R(\Gamma)+\lambda R^2(\Gamma)}$. This
``~interaction~''
lagrangian can be interpreted as first order General Relativity plus
a quadratic
correction in which $\rho$ plays the role of the ``~bare~''
cosmological constant and $\mu$ can be identified with the Planck mass.
Note that if we set $R(\Gamma)\equiv \phi^2$, where
$\phi$ is a scalar field, then $L_{int.}(R)$ takes the form of a
{\it Higgs potential}, and one may wonder about spontaneous symmetry
breaking of Weyl invariance. However, in spite of this formal
similarity, one should keep in mind that
Weyl invariance is broken explicitly, rather than spontaneously,
by the very presence of an interaction term, regardless of the
specific form of $L_{int.}(R)$.

Returning to the general case and to the formation of
structure, we suggest to interpret $L'_{int.}$ in equation (2.12) as
an {\it order parameter} for the geometric phases on the p--brane
manifold. The essential property which makes this interpretation
possible is that an analytic function has a {\it discrete}
number of zeros within its analyticity domain. Since the whole left
hand side of eq.\ (\ref{sette}) is an analytic function of $R$, there can be
only a discrete set of solutions, say $\{c_i\}$, such that
\begin{equation}
c_i L_{int.}^\prime (c_i)-{p+1\over 2}L_{int.}(c_i)=0
\ ,\qquad R=c_i\ .\label{nove}
\end{equation}
Hence, the conformal p--brane geometry admits two distinct
phases characterized by the ``~order parameter~''
$L_{int.}^\prime (c_i)=0$ , or, $L'_{int.}\ne 0$.
In the first instance , the scalar curvature $R=c_i$ is an extremal of
$L_{int}$ and eq.(2.14) implies $L_{int}(c_i)=0$ . When
$L'_{int}(c_i)\ne 0$, eq.(2.14) implies $L_{int}(c_i)\ne 0.$ We will
argue, next, that in correspondence of each of these cases there exists
a distinct geometric phase with a characteristic cellular structure on the
p--brane manifold.

\subsection{ Riemannian geometric phase}

	If the curvature extremizes the ``~potential~'', i.e.
$L_{int.}^\prime (c_i)=0$, then equation (\ref{nove})
requires $L_{int.}(c_i)=0$. Equation (\ref{sei}) is trivially
satisfied, and
the connection is no longer dynamically determined but can be freely
chosen. In this case,
eq.(\ref{cinque}) simplifies and becomes,
\begin{eqnarray}
&&\gamma^{mn}\left[
{1\over(p+1)}\gamma^{pq}\partial_p X^\mu \partial_q X_\mu
\right]^{\left(p+1\right)/2}+\nonumber\\
-&&\gamma^{mi}\gamma^{nj}\partial_i X^\rho\partial_j X_\rho
\left[{1\over(p+1)}\gamma^{pq}\partial_p X^\mu \partial_q X_\mu
\right]^{\left(p-1\right)/2}=0.\label{dieci}
\end{eqnarray}
{}From this equation it follows that the world volume metric can be
written as the induced metric times an arbitrary function of the
world coordinates,
\begin{equation}
\gamma_{mn}=\Omega(\xi)\partial_m X^\mu \partial_n X_\mu\ .
\label{dodici}
\end{equation}
Thus, this geometric phase corresponds to a Riemannian  background
geometry which is governed by the first order,
contracted, Einstein equation
$R=c_i$. Evidently, {\it for each $c_i$}, this equation
describes a spacetime of constant curvature (p--cell). Thus, barring
any degeneracy in the set of solutions $\{c_i\}$, one is led to the
conclusion that the dynamics of a p--brane induces a cellular
structure on the p--brane manifold. For $p=3$, each cell consists of
a three dimensional region separated from other cells by domain
walls and the over all structure resembles an ``~emulsion~''
\cite{lego}, or a ``~soap bubble froth~'' in which the dynamics of
each bubble is governed by matching conditions on the metrics of
neighboring cells.

Note, incidentally, that the contracted Einstein equation
$R=c_i$ represents a generalization of the basic equation of (1+1)--
dimensional gravity. As a matter of fact Eq.(\ref{dodici}) holds
true for any p--brane and is a
consequence of the Weyl invariance of the p--brane action. However,
while conformal invariance allows a common formal treatment of
strings and higher dimensional objects, the {\it role} played by
conformal invariance is distinctly unique in the case of strings.
For instance, equation
(\ref{dodici}) does {\it not} imply that the p--brane manifold is
conformally flat except in the string case, $p=1$, for which one can find a
coordinate transformation which maps the induced metric into a flat
metric. A necessary and sufficient condition for {\it conformal flatness}
of higher dimensional manifolds with $p+1\ge 4$ is that the Weyl tensor
vanishes.

\subsection{ Riemann--Cartan geometric phase}

If $L_{int.}^\prime (c_i)\ne 0$ then $\Gamma^m{}_{nr}$ becomes a
dynamical variable. In fact, eq.\ (\ref{sei}) gives
\begin{equation}
\nabla_a\left[\sqrt{-\gamma}\gamma^{mn}\right]=0\longrightarrow
\nabla_a\gamma^{mn}={\gamma^{mn}\over\sqrt{-\gamma}}\nabla_a\sqrt{-
\gamma} .
\label{tredici}
\end{equation}
But,
\begin{equation}
\nabla_a\sqrt{-\gamma}={1\over 2}\sqrt{-
\gamma}\left[\gamma^{mn}\partial_a
\gamma_{mn}-2\Gamma^m{}_{ma}\right] \ ,\label{quattordici}
\end{equation}
so that eq.\ (\ref{tredici}) can be written in the form
\begin{equation}
\nabla_a\gamma^{mn}={1\over 2}\gamma^{mn}\left[\gamma^{rs}\partial_a
\gamma_{rs}-2\Gamma^l{}_{la}\right]\ .
\label{unoquattro}
\end{equation}
To solve eq.\ (\ref{unoquattro}), we recall that a general affine
connection can always be written as the Christoffel symbol plus a term, say
$K^l{}_{mn}$, which behaves as a tensor under general
coordinate transformation
\begin{equation}
\Gamma^l{}_{mn}=\{{}_m{}^l{}_n\}+K^l{}_{mn}\ .
\label{ansatz}
\end{equation}
The Christoffel symbol $\{{}_m{}^l{}_n\}$ is a metric compatible
connection, so that the ansatz
(\ref{ansatz}), once inserted into eq.\ (\ref{unoquattro}), gives us
an equation for the tensor part $K^l{}_{mn}$ alone
\begin{equation}
(p-1)K^l{}_{lm}=0\ ,
\label{traccia}
\end{equation}
where we have used the identity $\displaystyle{
\{{}_m{}^l{}_l\}=(1/2)\gamma^{ab}
\partial_m \gamma_{ab} }$.
Eq.\ (\ref{traccia}) shows that, for any extended object different
from the string, the trace of $K^l{}_{mn}$ must vanish, so that
eq.\ (\ref{unoquattro}) for the ansatz \ (\ref{ansatz}) reduces to
\begin{equation}
\nabla_a\gamma^{mn}=0\Longrightarrow K^l{}_{pq}={1\over 2}\left(
T^l{}_{pq}+T_{pq}{}^l+T_{qp}{}^l\right)\ ,
\end{equation}
where $\displaystyle{T^l{}_{pq}=(1/2)\left(\Gamma^l{}_{pq}-
\Gamma^l{}_{qp}\right)}$ is the {\it torsion tensor}, and
$\Gamma^l{}_{pq}$ is identified with the {\it Riemann--Cartan
connection.}
This new geometric phase is also characterized by a cellular
structure, since the scalar curvature is still subject to the
constraint $R(\gamma,\Gamma)=c_i$. The novelty in this case is the
appearance of a cosmological constant with a cell--dependent value.
Indeed, in this geometric phase, eq.\ (\ref{cinque}) reduces to
the Einstein--Cartan field equation
\begin{equation}
R^{(mn)}(\Gamma) -{c_i\over p+1}\gamma^{mn}=
-{1\over L_{int.}^\prime (c_i) }T^{mn}(X)\ ,
\label{quindici}
\end{equation}
 where $-(1/ L_{int.}^\prime (c_i))$ plays the role of Newton's constant,
and $c_i/(p+1)$ acts as an effective cosmological constant in any
given cell on the p--brane manifold. It is interesting how Newton's constant
and the cosmological constant are related by the above formalism.
Evidently both originate from the set of
solutions $\{c_i\}$ of equation (2.14) (analyticity assumption). As
anticipated in the Introduction, it is this assumption that allows
us to bypass our ignorance of quantum gravitational effects: if a
generic p--cell has a linear dimension of the order of Planck's
length at the time of its nucleation, then the analyticity
assumption is tantamount to state that the quantum fluctuations in
the background metric are of the same order of magnitude as the
metric itself, which is the central consideration behind the
geometrodynamic idea of spacetime foam. Once the nucleation of
p--cells has taken place, the problem of their evolution is largely
a classical and tractable one \cite{spall}, and this is the point of
view advocated in  this paper.

	Finally, it should be noted that our formalism also provides
an insight into the question of the special status that strings hold
among p--branes: the point is that, for $p=1$, eq.(\ref{traccia}) is
satisfied by {\it any} $K^l{}_{lm}$. This means that the connection
$\Gamma^l{}_{qp}$ is defined up to an arbitrary vector field
$B_m\equiv -K^l{}_{lm}$. Accordingly, eq.(\ref{unoquattro}) becomes
\begin{equation}
\nabla_a\gamma^{mn}=\gamma^{mn}B_a\ .
\label{semimetric}
 \end{equation}
Eq.(\ref{semimetric}) is the {\it semi--metric} condition for the
Riemann--Weyl
connection \cite{vol}
\begin{equation}
\Gamma^l{}_{qp}=\{{}_q{}^l{}_p\}+{1\over 2}\left(\delta^l_p B_q +
\delta^l_q B_p-\gamma_{pq}B^l\right)
\end{equation}
where $B_p$ acts as the Weyl ``~gauge potential~'' associated with
volume--changing scalings.

	Thus, we conclude that the intrinsic
geometry on the world--sheet of a string is characterized by the pair
$(\gamma_{mn},\,B_p)$, while for a generic p--brane the geometrical objects are
the metric and a traceless torsion tensor. Furthermore,
the above results seem to be independent of any special length or
energy scale but seem suggestive enough to be given a cosmological
interpretation {\it at, or near the Planck scale}
in the physically interesting case in which the p--brane consists of
a spatial 3--dimensional manifold, $p=3$. In this case, the non--
minimal coupling term to the bag curvature
gives rise  to a ``~gravitational action~'' whose effect is to form
a cellular structure on the manifold. This structure is not static,
but a highly dynamical one which evokes, at least in our mind, a
vivid picture of the ground state of the primordial universe not
unlike the chaotic inflation scenario\cite{linde}. In the light of
the above results, the physical spacetime
can be pictured as a set of cells in which the geometry
is dynamically determined and not fixed at the outset. In
this scenario, extended objects (strings and membranes) may well
play a role comparable, or even alternative, to that of the Higgs field,
as the universe bootstraps itself into existence out of the primordial
spacetime foam. In this paper we have suggested that this structure
is a manifestation of the underlying multiphase geometry induced by the
very dynamics of p--branes encoded in the action (\ref{tre}). In
this interpretation, the cosmic vacuum
is a multi--phase system in a double sense: inside a cell there may
exist a
Riemannian or a Riemann--Cartan geometry; furthermore, for each type
of geometry,
curvature can attain different constant values labelled by $c_i$.
These parameters, in turn, determine the value and sign of the
energy density in each cell. Consequently, each cell may behave as a
blackhole, wormhole, inflationary bubble, etc.. The classical and
semiclassical evolution of any such cell has been discussed in
earlier papers \cite{spall}. Here, as a final note, we add that a
semi--classical description of the quantum
mechanical ground state, for such a
multi--domain system, is obtained by approximating the (euclidean)
Feynman integral with the sum over classical solutions. The non--minimal
interaction term in eq.\ (\ref{tre})  acts as
an {\it effective cosmological constant} once evaluated on a
classical solution. Thus, the cosmological constant enters the model
as a semi--classical dynamical variable and, therefore, is susceptible
of dynamical adjustments\cite{smail}. From this view point, the vanishing of
$L_{int.}(c_i)$ in the Riemannian phase is an attractive result.


\begin{references}
\bibitem{noi} A.Aurilia and E.Spallucci,
``~ The Role of Extended Objects in Particle Theory and in
Cosmology~''
Proceedings of the Trieste Conference on Super--Membranes and
Physics
in $2+1$ dimensions~'' Trieste, 17--21 June, 1989;
ed. M.J.Duff, C.N.Pope, E.Sezgin;  World Scientific, 1990.
\bibitem{cole} See, for instance, E.A.B.Cole,
Nuovo\ Cimento\ {\bf 1A}, 120, (1971).
\bibitem{linde} A.Linde,
``~Particle Physics and Inflationary Cosmology~''
(Harwood Academic, New York, 1990).
\bibitem{wheeler} J.A.Wheeler,
Ann.\ Phys.\ (NY)\ {\bf 2}, 604, (1957).
\bibitem{lego} A.Aurilia, G.Denardo, F.Legovini and E.Spallucci,
Nucl.\ Phys.\ {\bf B252}, 523 (1984).
\bibitem{vanz} A.Aurilia, E.Spallucci and I.Vanzetta,
Phys.\ Rev.\ {\bf D50}, 6490 (1994).
\bibitem{ht} P.S. Howe and R.W.Tucker,
J.\ Phys.\ {\bf A10}, L155, (1977).
\bibitem{suga} A.Sugamoto,
Nucl.\ Phys.\ {\bf B215}, 381, (1981).
\bibitem{Duff} M.S.Alves and J.Barcelos--Neto,
Europhys.\ Lett.\ {\bf 7}, 395, (1988).\\
M.J.Duff,
Class.\ Quantum\ Grav. {\bf 6}, 1577, (1989).
\bibitem{smail} A.Aurilia, A.Smailagic and
E.Spallucci, Class.\ Quantum\ Grav.\ {\bf 9}, 1883, (1992).
\bibitem{to} M.Ferraris, M.Francaviglia and I.Volovich,
``~Universality of Einstein equations in Palatini formalism~'',
University of Torino preprint,TO--JLL--P 1/92.
\bibitem{vol} I.V.Volovich,
Mod.\ Phys.\ Lett.\ {\bf A8}, 1827, (1993).
\bibitem{spall} A.Aurilia, M.Palmer and E.Spallucci,
Phys.\ Rev.\ {\bf D40}, 2511, (1989).\\
A.Aurilia, R.Balbinot and E.Spallucci,
Phys.\ Lett.\ {\bf B262} 222, (1991).
\end{references}
\end{document}